\documentclass{article}

\usepackage{microtype}
\usepackage{graphicx}
\usepackage{booktabs} 
\usepackage{enumitem}
\usepackage{mathtools}
\usepackage{multirow}
\usepackage{subfigure}
\usepackage{caption}
\usepackage{hyperref}

\usepackage[accepted]{mlsys2025}

\mlsystitlerunning{Staggered Batch Scheduling: Co-optimizing Time-to-First-Token and Throughput for High-Efficiency LLM Inference}

\begin{document}

\twocolumn[
\mlsystitle{Staggered Batch Scheduling: Co-optimizing Time-to-First-Token and Throughput for High-Efficiency LLM Inference}



\mlsyssetsymbol{equal}{*}

\begin{mlsysauthorlist}

\mlsysauthor{Jian Tian}{bd}
\mlsysauthor{Shuailong Li}{bd}
\mlsysauthor{Yang Cao}{bd}
\mlsysauthor{Wenbo Cui}{bd}
\mlsysauthor{Minghan Zhu}{bd}
\mlsysauthor{Wenkang Wu}{bd}
\mlsysauthor{Jianming Zhang}{bd}
\mlsysauthor{Yanpeng Wang}{bd}
\mlsysauthor{Zhiwen Xiao}{bd}
\mlsysauthor{Zhenyu Hou}{bd}
\mlsysauthor{Dou Shen}{bd}
\end{mlsysauthorlist}

\mlsysaffiliation{bd}{Baidu Inc. Beijing, China}

\mlsyscorrespondingauthor{Jian Tian}{tianjian01@baidu.com}

\mlsyskeywords{Machine Learning, MLSys}

\vskip 0.3in

\begin{abstract}
The evolution of Large Language Model (LLM) serving towards complex, distributed architectures—specifically the P/D-separated, large-scale DP+EP paradigm—introduces distinct scheduling challenges. Unlike traditional deployments where schedulers can treat instances as black boxes, DP+EP architectures exhibit high internal synchronization costs. We identify that immediate request dispatching in such systems leads to severe \textbf{in-engine queuing} and \textbf{parallelization bubbles}, degrading Time-to-First-Token (TTFT). To address this, we propose \textbf{Staggered Batch Scheduling (SBS)}, a mechanism that deliberately buffers requests to form optimal execution batches. This temporal decoupling eliminates internal queuing bubbles without compromising throughput. Furthermore, leveraging the scheduling window created by buffering, we introduce a \textbf{Load-Aware Global Allocation} strategy that balances computational load across DP units for both Prefill and Decode phases. Deployed on a production H800 cluster serving DeepSeek-V3, our system reduces TTFT by \textbf{30-40\%} and improves throughput by \textbf{15-20\%} compared to state-of-the-art immediate scheduling baselines.
\end{abstract}
]



\printAffiliationsAndNotice{}  

\section{Introduction}
\label{sec:intro}

The exponential scaling of Large Language Models (LLMs)—from hundreds of billions to trillions of parameters—has necessitated a fundamental shift in inference architectures. While Sparse Mixture-of-Experts (MoE) models have successfully reduced activation costs, they introduce significant memory access challenges. To bridge the gap between arithmetic intensity and memory bandwidth, advanced deployment paradigms like \textbf{DP+EP (Data Parallelism + Expert Parallelism)}, exemplified by the DeepSeek-V3 architecture, have emerged. Unlike traditional Tensor Parallelism (TP), DP+EP decouples the inference roles, utilizing massive parallel groups (e.g., EP size of 320 for Decode ) to handle high-concurrency workloads.

However, this architectural complexity exposes a critical inefficiency in current scheduling systems. \textbf{Traditional schedulers operate under a "continuous service" assumption}, dispatching requests immediately upon arrival. While effective for monolithic instances, this approach fails in P/D-separated DP+EP clusters. Our analysis reveals that the Prefill phase operates effectively as a \textbf{non-preemptive, discrete batch process}. Immediate dispatching ignores the "busy state" of the engine, forcing requests to queue internally within the inference instance (Device-side Queuing). This results in \textbf{Head-of-Line (HOL) blocking}, where idle resources in other instances remain unutilized while requests wait in a saturated engine’s private queue.

To address this, we propose \textbf{Staggered Batch Scheduling (SBS)}, a novel paradigm tailored for large-scale DP+EP clusters. By introducing a precise, adaptive waiting window, SBS aggregates requests into optimal batches before dispatch. This counter-intuitive strategy—\textbf{waiting to accelerate}—effectively eliminates device-side queuing and minimizes Time to First Token (TTFT). Furthermore, the buffering window empowers the scheduler with a global view, enabling a \textbf{Load-Aware Global Allocation} mechanism critical for resolving the intrinsic load imbalance across DP units in both the Prefill and Decode phases.

Specifically, our contributions are:
\begin{itemize}[itemsep=0pt,topsep=0pt,parsep=0pt]
    \item \textbf{Architectural Characterization:} We identify the "discrete batching" nature of the Prefill phase in DP+EP systems and demonstrate formally why immediate dispatching is suboptimal for TTFT.
    \item \textbf{Staggered Batch Scheduling (SBS):} We propose an adaptive, interval-based scheduling algorithm that eliminates device-side queuing latencies. This reduces TTFT by 30-40\% in production environments without sacrificing throughput.
    \item \textbf{Load-Aware Global Allocation:} Exploiting the global view provided by the batching window, we introduce a unified allocation strategy for both Prefill and Decode phases. This optimization mitigates phase-specific load imbalances, boosting Prefill throughput by 12.9\%–22.8\% and enhancing Decode throughput by $\sim$15\%.
\end{itemize}
\section{Related Work}
\label{sec:related-work}

The performance optimization of large language model (LLM) inference systems critically depends on the deep co-design of scheduling strategies and system architecture. Inherent heterogeneity among inference phases renders traditional load-balancing strategies suboptimal. As the core of distributed inference architectures, the scheduling subsystem must dynamically balance TTFT(Time-To-First-Token), TPOT(Time-Per-Output-Token), and overall system throughput. At the same time, It must also support distributed inference services at scale, managing models with hundreds of billions of parameters and highly variable workloads.

\textbf{PD disaggregation architectures.} PD disaggregation serves as a foundational architecture in distributed LLM inference, which decouples the compute-intensive prefill stage from the memory-intensive decode stage by assigning them to dedicated instance pools. Splitwise \cite{patel2024splitwise} pioneered this approach with a three-tier pool design—comprising a prefill pool, a decode pool, and a hybrid pool—and achieved up to 1.76× higher throughput and 15\% lower power consumption under the same cost by transferring the KV cache over the network. Subsequent studies have further refined this architecture. For example, chunked prefill was introduced to alleviate bottlenecks in mixed-length request processing \cite{hu2024inference,agrawal2024taming}; instance flipping enabled dynamic role-switching between prefill and decode instances \cite{zhong2024distserve}; and a shared memory pool reduced redundant computation in multi-turn dialogues through unified KV cache management \cite{hu2024memserve}.

\textbf{Scheduling under Hybrid Parallelism}. In parallel to disaggregation, another major architectural direction exploits hybrid parallelism to support ever-larger models. Scaling model size has been shown to enhance model capabilities \cite{brown2020language,touvron2023llama,hoffmann2022training}. Mixture-of-Experts (MoE)  architectures further pushed model scale boundaries \cite{dai2024deepseekmoe,rajbhandari2022deepspeed,zoph2022st}. The DP+EP (Data Parallelism + Expert Parallelism) hybrid architecture became a key breakthrough \cite{sglang2024blog}, keeping attention modules under data parallelism while employing expert parallelism for MoE layers. However, while DP+EP reduces memory footprint per device, it introduces scheduling challenges: All-to-All communication amplifies single-device failure impact, and load imbalance constrains throughput.
To address these, scheduling strategies focused on system efficiency and locality optimization have been proposed. The Locality-aware Fair Scheduler \cite{cao2025locality} introduced the Deficit Longest Prefix Matching (DLPM) algorithm, improving single-machine throughput by 2.87x. Its distributed variant, D²LPM, uses a decentralized architecture with a global radix tree and deficit counters, reducing P99 latency by 7.18x. However, D²LPM lacks responsiveness to runtime workload dynamics. To solve dynamic imbalance, Llumnix \cite{sun2024llumnix} introduced cross-instance fine-grained request rescheduling with a near-zero-overhead KV migration mechanism, achieving 6.4x TTFT improvement, 12.1x P99 latency improvement, and up to 36\% cost savings.

\textbf{Prefix Cache-Aware Scheduling.} In dialogue, RAG, and multi-tenant scenarios, many requests share common prefixes such as conversation history or system prompts, resulting in redundant KV cache computation. Preble \cite{srivatsa2024preble} systematically optimized distributed prefix cache sharing by introducing a prefix tree and the E² scheduling algorithm, which balances reuse of existing caches with exploration of new sharing opportunities. Its global-local two-tier scheduler reduced average latency by 1.5–14.5× under shared-prefix workloads. SGL-Router \cite{zheng2024sglang} employed an approximate prefix tree and cache affinity score to enable communication-free routing, improving throughput by 1.9× and cache hit rate by 3.8×. Together, these works mark a shift from simple load balancing to intelligent, prefix-aware routing.
\section{Analysis}
\label{sec:analysis}

\subsection{Scheduling Granularity: From Monolithic Instances to DP-Attention Units}
\label{sec:analysis_schedule_unit}

In traditional centralized deployment paradigms, a monolithic inference instance serves as the \textbf{atomic scheduling unit}. Since a single instance handles both Prefill and Decode phases autonomously, schedulers typically map requests 1:1 to instances. However, the evolution toward \textbf{P/D disaggregated architectures} necessitates decoupling these roles: dedicated pools handle the compute-bound Prefill phase (one-time processing) and the memory-bound Decode phase (autoregressive generation) separately.

This granularity is further refined in large-scale \textbf{DP+EP (Data Parallelism + Expert Parallelism)} architectures. For example, in the DeepSeek-V3 deployment, a single Prefill instance operates with distinct parallel strategies for its components (e.g., Expert weights are shared across ranks, while Attention layers are replicated). Specifically, in a 32-GPU setup configured with Tensor Parallelism ($TP=4$), the Data Parallelism degree reaches $DP=8$. Consequently, the finest-grained scheduling unit shifts from the "whole instance" to the \textbf{individual DP-Attention processing unit}. This architectural shift implies that a scheduler must manage resources at the sub-instance level to fully exploit hardware parallelism.

\begin{figure}[h]
    \centering
    \subfigure[\textbf{Traditional Monolithic Architecture:} The scheduler treats the entire inference instance as a single atomic unit. This coarse granularity masks internal resource fragmentation.]{
        \includegraphics[width=0.4\textwidth]{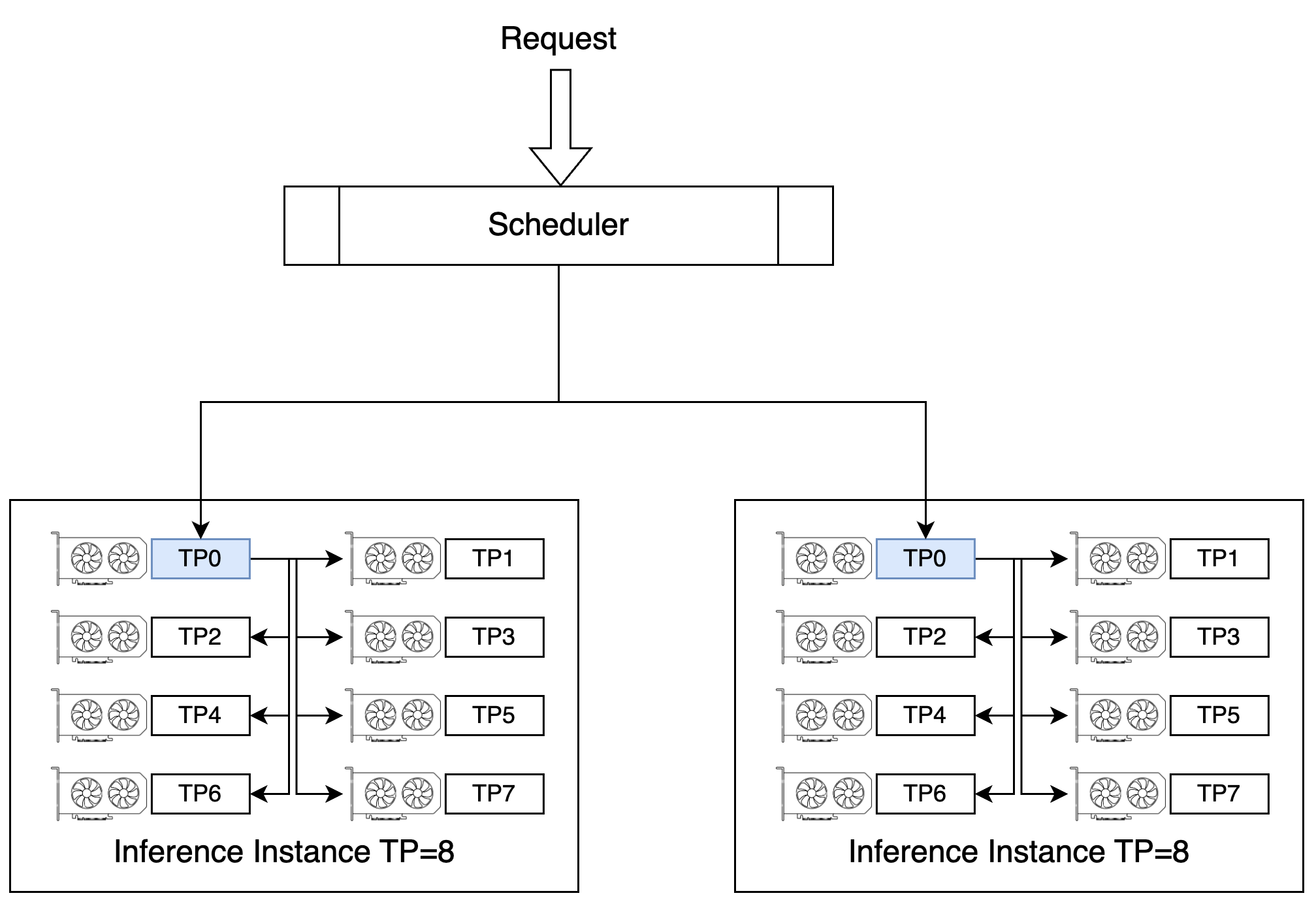}
        \label{fig:schedule_unit_1}
    } 
    \subfigure[\textbf{DP+EP Disaggregated Architecture:} In large-scale deployments, the scheduling unit shifts to the fine-grained \textbf{DP-Attention Group}. This exposes the need for sub-instance resource management to handle the complex mapping between Data Parallelism (DP) and Expert Parallelism (EP).]{
       \includegraphics[width=0.4\textwidth]{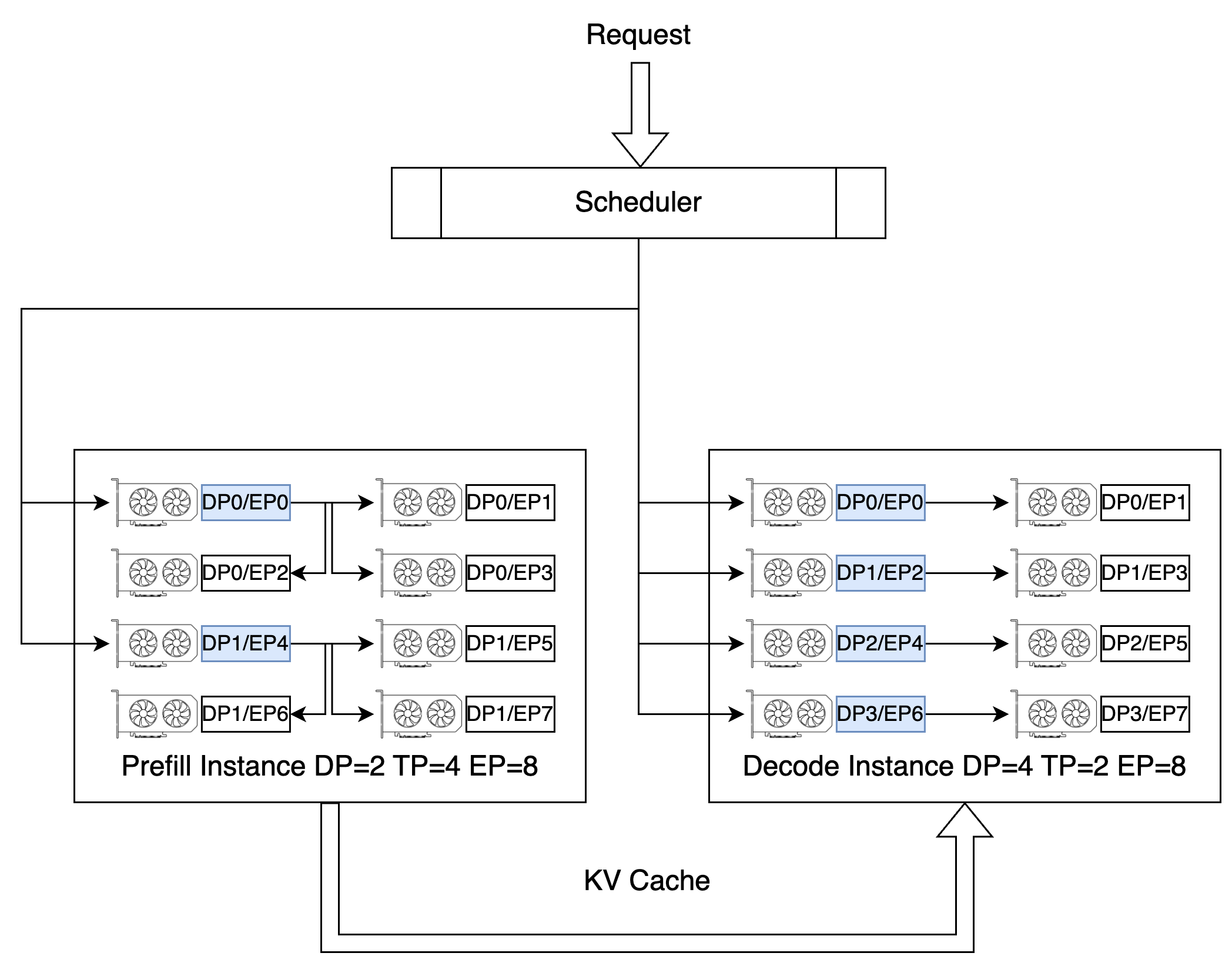}
       \label{fig:schedule_unit_2}
    }
    \caption{Evolution of Scheduling Granularity.}
    \label{fig:schedule_unit}
\end{figure}

\subsection{Queuing Dynamics and Head-of-Line Blocking}
\label{sec:analysis_queue_phenomena}

In traditional deployments, inference instances are often modeled as continuous service queues (approximating M/M/1 systems), where immediate dispatch minimizes latency. While this holds for the Decode phase due to its short Time-Per-Output-Token (TPOT), the Prefill phase in DP+EP architectures exhibits fundamentally different characteristics:
\begin{enumerate}[itemsep=2pt,topsep=0pt,parsep=0pt]
\item \textbf{Discrete Gated Service:} A Prefill instance operates as a non-preemptive, discrete batch processor. Once a forward pass begins, the engine enters a "locked" state and cannot accept new inputs until the current batch completes.
\item \textbf{Batch-Insensitive Latency:} Within capacity limits, the execution time for a batch is largely dominated by the longest sequence and synchronization overhead, rather than the batch size itself.
\end{enumerate}

Under these conditions, an \textbf{Immediate Dispatch} strategy becomes counter-productive. It blindly pushes requests to busy instances, causing them to accumulate in the engine's internal input queue. This results in \textbf{Head-of-Line (HOL) blocking}, where a request effectively waits for the full duration of the current batch execution ($T$) inside the device, unobservable and unmanageable by the scheduler.

We analyze this using a simplified model: assuming uniformly arriving requests and $N$ inference instances with processing time $T$.
\begin{itemize}[itemsep=2pt,topsep=0pt,parsep=0pt]
    \item \textbf{Immediate Dispatch:} Requests are assigned immediately. The expected waiting time effectively occurs inside the engine (Device-side Queuing), averaging $T/2$ regardless of cluster size $N$.
    \item \textbf{Staggered Batch Dispatch:} The scheduler buffers requests for a short interval ($T/N$) to form a batch. This shifts the waiting time to the Scheduler-side Queue.  
\end{itemize}

\begin{figure}[h]
    \centering
    \subfigure[\textbf{Immediate Dispatch (Baseline):} Requests are assigned instantly upon arrival. Because the engine is non-preemptive, requests accumulate in the \textbf{Device-side Queue} (red blocks), causing \textbf{Head-of-Line (HOL) blocking} and high latency.]{
        \includegraphics[width=0.45\textwidth]{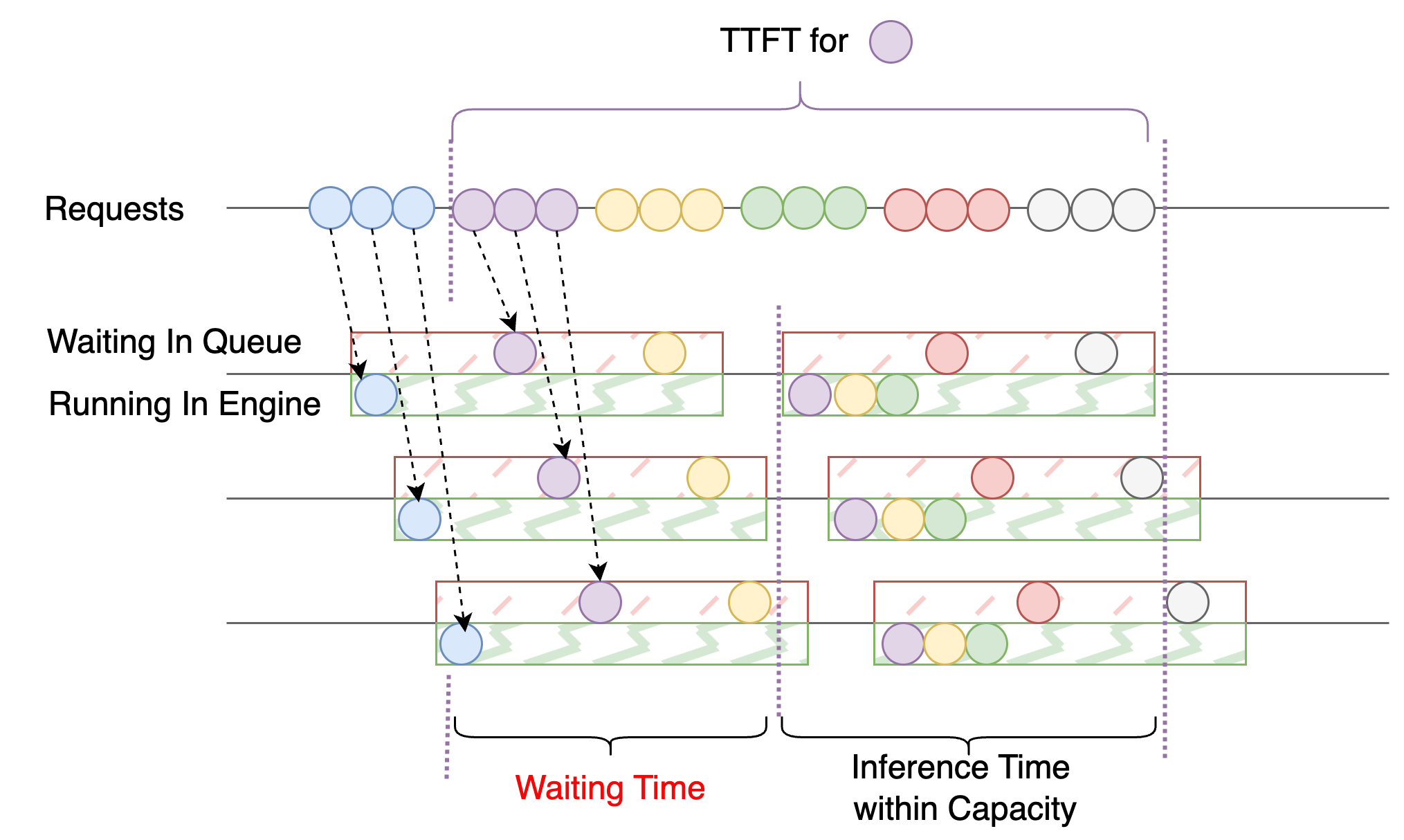}
        \label{fig:queue_1}
    } 
    \subfigure[\textbf{Staggered Batched Scheduler(Ours)}: Incoming requests (represented by colored circles) are grouped into batches before being submitted to an available engine. When new requests (e.g., purple, yellow) arrive, they are directed to an engine that is ready or will soon be ready. This  "staggered batch" scheduling significantly reduces "Waiting Time" by preventing requests from accumulating in a queue. ]{
       \includegraphics[width=0.45\textwidth]{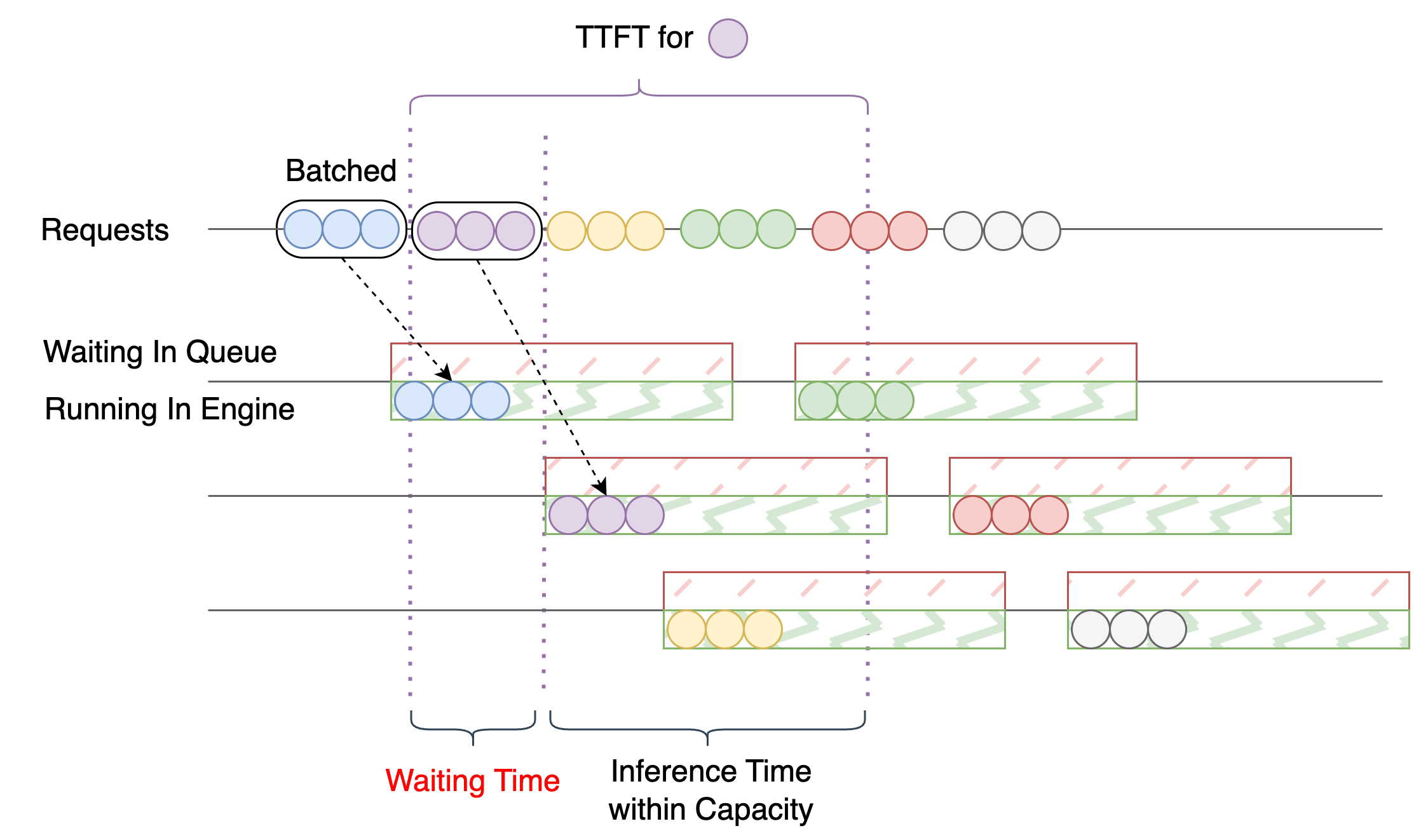}
       \label{fig:queue_2}
    }
    \caption{Impact of Dispatch Strategy on Queuing Dynamics.}
    \label{fig:queue}
\end{figure}

By enforcing this staggered interval, the expected total queuing delay drops from $T/2$ to $T/2N$. For large-scale clusters (e.g., $N > 10$), this theoretically yields an order-of-magnitude reduction in waiting latency. As illustrated in Figure \ref{fig:queue_2}, The scheduler buffers requests in a Scheduler-side Queue to form optimal batches. This eliminates internal HOL blocking, ensuring that requests enter the engine only when resources are ready, thereby minimizing total wait time.

We formalize the inference dynamics using Queuing Theory. Ideally, an $N$-instance cluster should function as an \textbf{M/D/S} system (Markovian arrival, Deterministic service, $S=N$ servers), maximizing resource pooling. However, under \textbf{Immediate Dispatch}, the scheduler prematurely binds requests to specific instances, effectively degrading the system into $N$ isolated \textbf{Gated M/D/1} queues.

Unlike the Decode phase which approximates a continuous flow, the Prefill phase's Gated M/D/1 nature implies that service cannot commence until a batch is fully formed and the pipeline clears. Consequently, a request assigned to a busy instance suffers from \textbf{Head-of-Line blocking}, with an expected waiting time dominated by the single instance's cycle time ($T/2$).

By enforcing a synchronized scheduling interval, our strategy effectively virtualizes the cluster back into a unified \textbf{M/D/S} system. This shifts the queue from the fragmented device-side (unmanageable) to the global scheduler-side (manageable), reducing the expected latency from $T/2$ to $T/(2N)$.

\subsection{Synchronization Overhead and The Batching Opportunity}
\label{sec:analysis_load_balance}

While traditional load balancing (e.g., Round-Robin or Least-Outstanding-Requests) suffices for monolithic instances, it fails in distributed DP+EP systems due to strict synchronization requirements.

In a DP+EP architecture, the Mixture-of-Experts (MoE) layers require \textbf{All-to-All communication}, meaning the MLP computation for a layer cannot proceed until all DP-Attention workers complete their attention mechanism. Consequently, the system throughput is bound by the slowest DP worker (the \textbf{straggler effect}). If workloads are unevenly distributed among DP units, faster units must idle-wait for the straggler, leading to significant synchronization overhead and resource wastage.

\begin{figure}[h]
    \centering
    \includegraphics[width=0.45\textwidth]{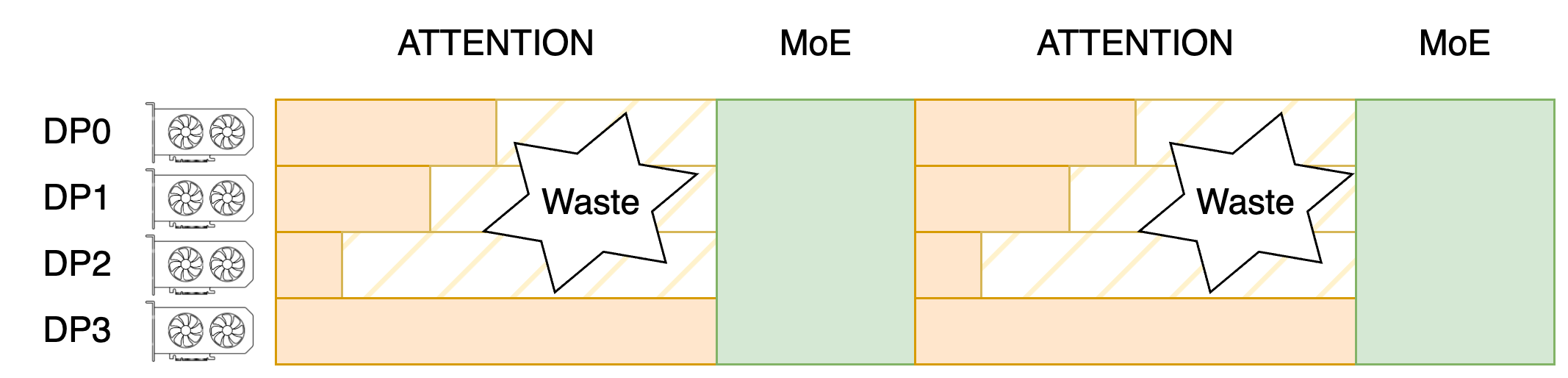}
    \caption{\textbf{Synchronization Overhead under Immediate Dispatch.} Due to the strict synchronization barrier in DP+EP architectures, the system throughput is bottlenecked by the slowest DP unit (Straggler). Greedy assignment leads to load imbalance, resulting in significant Parallelization Bubbles (marked as "Waste") where faster DPs idle-wait for stragglers .}
    \label{fig:load_balance_1}
\end{figure}

Achieving ideal load balance across DP units requires a global view of pending requests. Immediate dispatch strategies, being inherently greedy, assign requests sequentially based on the instantaneous system state, often resulting in suboptimal local decisions.

However, the \textbf{Staggered Batch Scheduling} strategy proposed in Section \ref{sec:analysis_queue_phenomena} introduces a critical side-effect: the \textbf{Batching Window}. By buffering requests to optimize latency, the scheduler inadvertently gains a temporal window to observe a pool of pending requests. This transforms the scheduling problem from a sequential, online decision process into a \textbf{mini-batch global optimization problem}.

This window enables us to apply sophisticated allocation algorithms (e.g., sorting and bin-packing) to distribute the workload evenly across DP-Attention units. Thus, the "waiting period"—initially introduced to reduce queuing latency—simultaneously serves as the enabler for fine-grained, sync-aware load balancing, co-optimizing both TTFT and throughput.

\begin{figure}[h]
    \centering
    \includegraphics[width=0.45\textwidth]{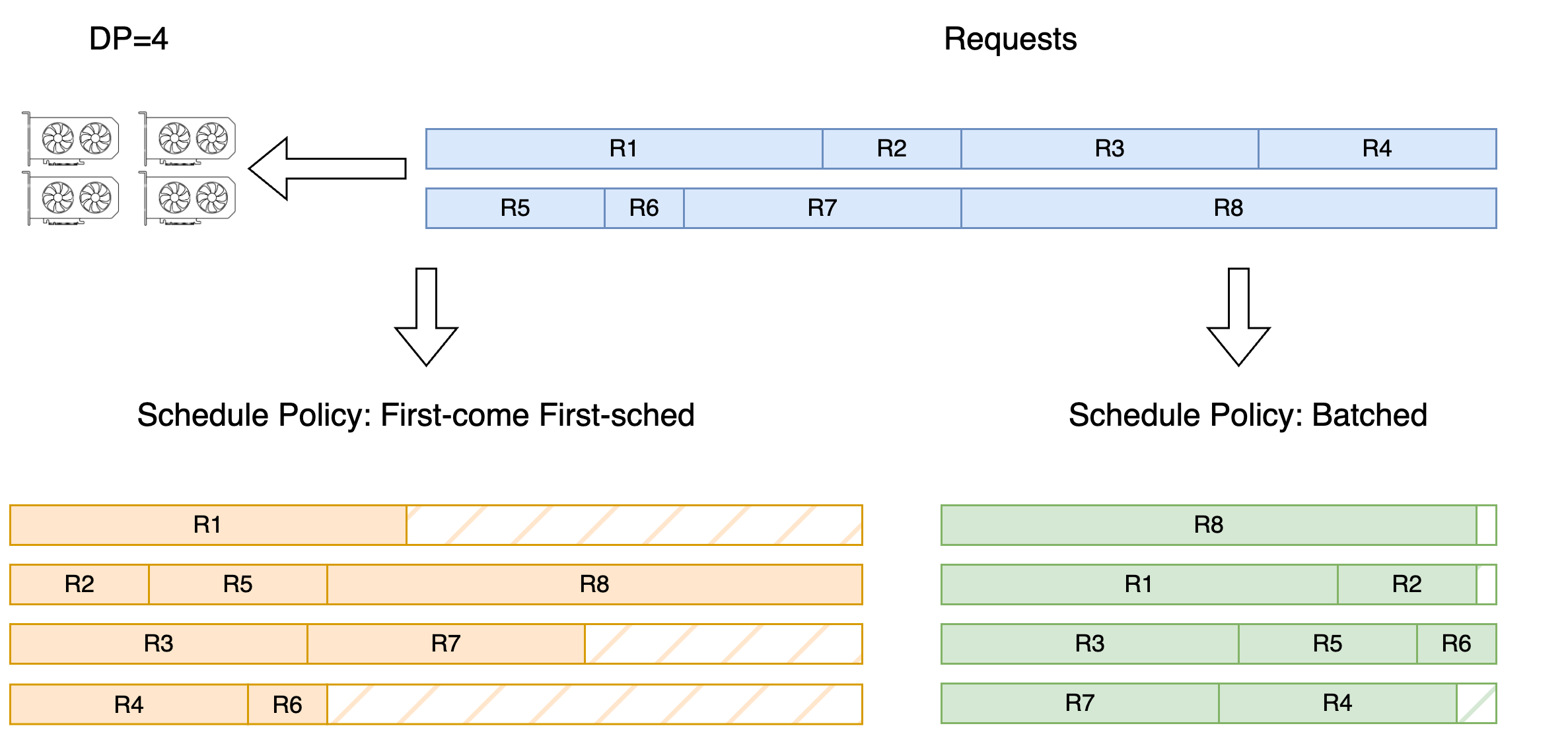}
    \caption{\textbf{Mitigation of Straggler Effect via Batched Bin-Packing.} By buffering requests to form a batch, the \textbf{Staggered Batch Scheduler} gains a global view to apply "Water-Filling" allocation (Algorithm \ref{alg:pbaa}). This ensures uniform workload distribution across DP units, filling the bubbles seen in Figure \ref{fig:load_balance_1} and maximizing effective compute utilization .}
    \label{fig:load_balance_2}
\end{figure}

\section{Design and Implementation}
\label{sec:design}
\begin{figure}[h]
    \centering
    \includegraphics[width=0.45\textwidth]{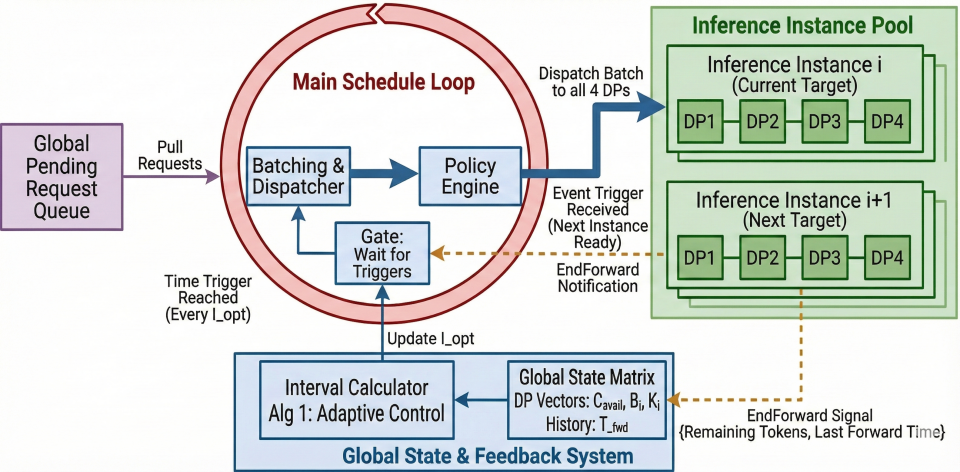}
    \caption{\textbf{System Architecture of the Staggered Batch Scheduler (SBS).} The system is centered around a Main Schedule Loop that governs request dispatching. (1) Inference Instances, each consisting of multiple Data Parallel (DP) units, execute forward passes. Upon completion of a pass, they asynchronously send an EndForward Signal containing payload statistics (remaining token count and execution time) back to the scheduler. (2) The Global State \& Feedback System acts as the source of truth, maintaining the Global State Matrix ($\langle C_{avail}, B_i, K_i \rangle$) updated by instance feedback, and dynamically calculating the optimal interval ($I_{opt}$) via Algorithm \ref{alg:adaptive_interval}. (3) The Schedule Loop waits for a dual trigger condition: the elapse of the calculated interval $I_{opt}$, AND the receipt of an EndForward notification from the next target instance. Once triggered, the scheduler batches pending requests and dispatches them to all DPs of the selected instance via the Policy Engine(Algorithm \ref{alg:pbaa} \& \ref{alg:iqr_schedule}), initiating the next cycle.}
    \label{fig:scheduler_architecture}
\end{figure}

Our proposed \textbf{Staggered Batch Scheduler (SBS)} operates around a unified closed-loop feedback control mechanism. As illustrated in Figure \ref{fig:scheduler_architecture}, the system organizes the end-to-end inference workflow through three tightly coordinated planes: the Control Plane (main scheduling loop), the State Plane (global state and feedback system), and the Resource Plane (inference instance pool).

\subsection{Adaptive Scheduling Interval \& Synchronization}
\label{sec:scheduling_interval}

\subsubsection{Throughput-Adaptive Interval Estimation}
The efficacy of SBS hinges on the precise alignment of the dispatch interval with the cluster's aggregate service rate. A static interval is insufficient for online environments characterized by high traffic volatility ($>100\%$ peak-to-trough variance).

We formulate the interval estimation as a dynamic control problem to balance system throughput against TTFT. The optimal scheduling interval, $I_{opt}$, is dynamically derived to match the request arrival rate with the cluster's aggregate processing capacity. We define this relationship as:$$I_{opt} = \frac{\bar{T}_{fwd} + L_{net}}{N_{active}}$$where $\bar{T}_{fwd}$ represents the moving average of the forward pass execution time, $L_{net}$ denotes the network latency for request distribution, and $N_{active}$ is the current number of healthy inference instances.

To ensure stability against transient jitter, the scheduler maintains a sliding window of reported execution times ($W_{stats}$), applying a moving average filter to update $\bar{T}_{fwd}$. This feedback loop allows the scheduler to converge rapidly to the optimal cadence following auto-scaling events or workload shifts.

During system initialization, due to the lack of online runtime data, a default value derived from offline stress testing is used as the initial value for $\bar{T}_{fwd}$. As the system continues to operate, the scheduling interval rapidly converges to the optimal value for the current environment through the aforementioned adaptive update mechanism.

\begin{algorithm}[h]
\caption{Throughput-Adaptive Interval Control Loop}
\label{alg:adaptive_interval}
\begin{algorithmic}[1]
\REQUIRE
    $W_{size}$: Maximum size of the sliding window;
    $L_{net}$: Estimated network overhead latency;
    $T_{default}$: Initial fallback execution time.
\ENSURE Optimal scheduling interval $I_{opt}$.

\STATE \textbf{Global State:}
\STATE \quad $\mathcal{W} \gets \emptyset$ \COMMENT{Sliding window for execution times}
\STATE \quad $\bar{T}_{fwd} \gets T_{default}$ \COMMENT{Smoothed forward time}
\STATE \quad $N_{active} \gets \text{GetActiveInstances}()$

\STATE
\FUNCTION{RecomputeInterval}
    \IF{$N_{active} > 0$}
        \STATE $I_{opt} \gets (\bar{T}_{fwd} + L_{net}) / N_{active}$
        \STATE \textbf{Update} System Timer with $I_{opt}$
    \ENDIF
\ENDFUNCTION
\STATE

\STATE \COMMENT{Triggered upon receiving 'EndForward' signal from any instance}
\FUNCTION{OnEndForward({$t_{measured}$})}
    \STATE $\mathcal{W}.\text{Enqueue}(t_{measured})$
    \IF{$\text{Size}(\mathcal{W}) > W_{size}$}
        \STATE $\mathcal{W}.\text{Dequeue}()$ \COMMENT{Evict oldest sample}
    \ENDIF
    \STATE $\bar{T}_{fwd} \gets \text{Mean}(\mathcal{W})$ \COMMENT{Apply moving average filter}
    \STATE RecomputeInterval
\ENDFUNCTION
\STATE
\STATE \COMMENT{Triggered by Auto-scaler or Health-check events}
\FUNCTION{OnTopologyChange($N_{new}$)}
    \STATE $N_{active} \gets N_{new}$
    \STATE RecomputeInterval \COMMENT{Immediate adaptation to capacity shift}
\ENDFUNCTION
\end{algorithmic}
\end{algorithm}

\subsubsection{Robust State Synchronization Protocol}

Relying solely on estimated intervals for request dispatching is susceptible to synchronization drift caused by transient workload fluctuations and variable execution times. Although the adaptive interval algorithm converges rapidly, transient estimation errors may still degrade performance or, in extreme cases, lead to cluster-wide deadlocks. To guarantee system liveness and maximize resource utilization, we implement a \textbf{Multi-tier State Synchronization Protocol} that supplements the interval estimation. This protocol employs a triple-check mechanism to ensure the accuracy and robustness of readiness judgments:

\begin{enumerate}[itemsep=2pt,topsep=0pt,parsep=0pt]
    \item \textbf{Quiescence Polling (Initialization Path)}: The scheduler continuously monitors the task depth of instance queues. A zero-task state acts as an immediate trigger for readiness. This mechanism is particularly effective for minimizing latency during system initialization (cold starts) and for rapid recovery following the completion of a batch.
    \item \textbf{Asynchronous Completion Signaling (Fast Path)}: Serving as the standard cooperative mechanism, instances proactively push an EndForward event to the scheduler upon completing a forward pass. This event-driven approach serves as the primary, low-latency trigger for updating load status and signaling instance readiness.
    \item \textbf{Liveness Watchdog (Safety Path)}: To tolerate network partitions or silent instance faults where EndForward events are lost, a watchdog timer is armed upon dispatch. The timeout threshold is set to $T_{\text{timeout}} = 5 \times \bar{T}$. Expiration of this timer forces a state reset, preventing distributed deadlocks and ensuring system liveness.
\end{enumerate}

This three-pronged strategy provides a comprehensive safeguard against abnormal scenarios. Notably, the watchdog mechanism enables graceful degradation: under worst-case conditions (e.g., complete loss of contact with an instance), the system automatically reverts to a fixed-interval batch processing mode. This design effectively prevents global blocking, guarantees fundamental service availability, and demonstrates high system robustness.

\subsection{Batched Prefill Dispatching  }
\label{sec:batched_prefill}

In large-scale DP+EP architectures, the strict \textbf{synchronization barrier} across Data Parallel (DP) units implies that the end-to-end latency of a batch is dictated by the heaviest workload (the straggler). While inference engines typically employ chunked prefill to decompose long sequences into manageable execution units, traditional schedulers remain agnostic to this fine-grained decomposition. They allocate resources based on the coarse-grained total request length rather than the actual chunk capacity. This \textbf{granularity mismatch} between scheduler perception and execution reality leads to severe load imbalance and resource fragmentation. To resolve this, we establish a precise \textbf{Dynamic Capacity Model} and design a two-level allocation algorithm that proactively balances load at the chunk level.

\begin{algorithm}[H]
\caption{Prioritized Batch Allocation Algorithm (PBAA)}
\label{alg:pbaa}
\begin{algorithmic}[1]
\REQUIRE
    $\mathcal{Q}_{pending}$: Unassigned requests from previous cycles;
    $\mathcal{Q}_{new}$: Newly arrived requests;
    $\mathcal{D}$: Set of DP units with available capacity $C_{avail}^{(d)}$;
    $N_{limit}$: Maximum tolerable waiting cycles
\ENSURE Assignment mapping $\mathcal{M}$.

\FUNCTION{GreedyDispatch($\mathcal{Q}$)}
    \STATE Sort $\mathcal{Q}$ by length $\mathcal{L}(r)$ in descending order \COMMENT{Reduce fragmentation}
    \FOR{$r \in \mathcal{Q}$}
        \STATE \textbf{define} $\text{Capacity}(r, d)$:
        \STATE \quad \textit{Basic Mode:} $C_{avail}^{(d)} - \mathcal{L}(r)$
        \STATE \quad \textit{Cache-Aware:} $C_{avail}^{(d)} - (\mathcal{L}(r) - \mathcal{L}_{hit}(r, d))$
        \STATE $d^* \gets \arg\max_{d \in \mathcal{D}} \text{Capacity}(r, d)$ \COMMENT{Select optimal DP}
        
        \IF{$C_{avail}^{(d^*)} > 0 $}
            \STATE $\mathcal{M} \gets \mathcal{M} \cup \{r \to d^*\}$
            \STATE $C_{avail}^{(d^*)} \gets Capacity(r, d^*)$
        \ELSE
            \STATE $\mathcal{Q}_{next} \gets \mathcal{Q}_{next} \cup \{r\}$
        \ENDIF
    \ENDFOR
\ENDFUNCTION

\STATE
\STATE $\mathcal{M} \gets \emptyset, \mathcal{Q}_{next} \gets \emptyset$

\STATE \COMMENT{Phase 1: Prioritize Legacy}
\STATE GreedyDispatch($\mathcal{Q}_{pending}$) 
\STATE
\STATE  \COMMENT{Phase 2: Assign New Arrivals}
\STATE GreedyDispatch({$\mathcal{Q}_{new}$}) 

\STATE
\STATE \COMMENT{Phase 3: Overload detection}
\FOR{$r \in \mathcal{Q}_{next}$}
    \STATE $r.wait \gets r.wait + 1$
    \IF{$r.wait > N_{limit}$}
        \STATE \textbf{Trigger} FlowControl(Throttle/Reject)
    \ENDIF
\ENDFOR
\STATE \bf{return} $\mathcal{M}, \mathcal{Q}_{next}$
\end{algorithmic}
\end{algorithm}

\subsubsection{Fine-Grained DP Capacity Modeling}
The scheduler maintains a real-time state vector for every DP unit $d_i$ in the cluster. We define the \textbf{Real-time Available Capacity} ($C_{avail}^{(i)}$) as:$$C_{avail}^{(i)} = C_{chunk} - U_{flight}^{(i)} - R_{queued}^{(i)}$$Here, $C_{chunk}$ is the hardware-constrained maximum token capacity per forward pass, $U_{flight}^{(i)}$ represents tokens in transit (dispatched but unacknowledged), and $R_{queued}^{(i)}$ denotes the backlog currently buffered on the device. This model provides a precise view of "dispatchable headroom," enabling the scheduler to fill bubbles in the parallel pipeline.

\subsubsection{Capacity-Constrained Greedy Allocation}

We propose the \textbf{Prioritized Batch Allocation Algorithm (PBAA)} to map a buffered batch of requests $\mathcal{Q}$ to available DP units $\mathcal{D}$. As detailed in Algorithm~\ref{alg:pbaa}, the process operates in three phases:
\begin{enumerate}[itemsep=2pt,topsep=0pt,parsep=0pt]
\item \textbf{Starvation Prevention:} Pending requests from previous cycles are prioritized to strictly enforce First-Come-First-Serve (FCFS) fairness.
\item \textbf{Straggler-Aware Bin Packing:} We employ a greedy heuristic that assigns the most computationally intensive requests (longest sequences) to the DP unit with the highest available capacity $C_{avail}$. $$d_{target} = \arg\max_{d \in \mathcal{D}} (C_{avail}^{(d)})$$This "Water-Filling" strategy proactively balances the load before execution begins, minimizing intra-layer synchronization wait times.
\item \textbf{Overload Protection:} If a request fails allocation for $N$ consecutive cycles, it triggers a flow control mechanism (e.g., throttling) to prevent system-wide saturation.
\end{enumerate}

\textbf{Optimization for Context Caching:} In cache-enabled scenarios, the objective function shifts from raw capacity to \textbf{effective computational cost}. The selection criterion is refined to maximize the cache hit rate:$$d_{target} = \arg\max_{d \in \mathcal{D}} [C_{avail}^{(d)} - (\text{Len}(r) - \text{Len}_{hit}(r, d))]$$This directs requests to DPs retaining relevant KV caches, significantly reducing redundant attention computation.

\subsection{Dual-Objective Batched Decode Scheduling}
\label{sec:batched_decode}

\subsubsection{The Coupled Load Imbalance Challenge}
Unlike Prefill, the Decode phase is autoregressive, creating a unique \textbf{Coupled Load Imbalance} problem:

\begin{itemize}[itemsep=2pt,topsep=0pt,parsep=0pt]
    \item \textbf{Memory Imbalance:} The heavy-tailed distribution of sequence lengths can exhaust KV cache memory on specific DP units (stragglers).
    \item \textbf{Batch-Size Imbalance:} Uneven request counts lead to low GPU utilization and communication inefficiencies.
\end{itemize}
Existing strategies that optimize only one dimension often exacerbate the other. We propose a coordinated strategy that jointly optimizes both dimensions using the global view provided by the batching window.

\begin{algorithm}[H]
\caption{IQR-Aware Lexicographical Decode Scheduling}
\label{alg:iqr_schedule}
\begin{algorithmic}[1]
\REQUIRE
    $\mathcal{R}$: Batch of decode requests to be scheduled;
    $\mathcal{N}$: Set of decode DP units, each with state $\langle B_i, K_i \rangle$;
    $k$: IQR multiplier threshold (typically 1.5).
\ENSURE Updated DP units states and assignment mapping.

\STATE \textbf{function} $\text{LexCompare}(i, j)$:
\STATE \quad \textbf{return} $(B_i < B_j) \textbf{ or } (B_i = B_j \textbf{ and } K_i < K_j)$

\STATE
\FUNCTION{ScheduleBatch({$\mathcal{R}, \mathcal{N}$})}
    \STATE Sort $\mathcal{R}$ by total sequence length in descending order \COMMENT{Fill-the-valley strategy}
    
    \FOR{each request $r \in \mathcal{R}$}
        \STATE \textbf{Step 1: Outlier Detection (Masking)}
        \STATE $\mathbf{K} \gets \{ K_n \mid n \in \mathcal{N} \}$ \COMMENT{Snapshot of current KV loads}
        \STATE $Q_1, Q_3 \gets \text{Percentile}(\mathbf{K}, 25), \text{Percentile}(\mathbf{K}, 75)$
        \STATE $Th_{outlier} \gets Q_3 + k \cdot (Q_3 - Q_1)$
        
        \STATE $\mathcal{N}_{safe} \gets \{ n \in \mathcal{N} \mid K_n \le Th_{outlier} \}$
        \IF{$\mathcal{N}_{safe} = \emptyset$} 
            \STATE \COMMENT{Fallback if all DP units saturated}
            \STATE $\mathcal{N}_{safe} \gets \mathcal{N}$
        \ENDIF

        \STATE \textbf{Step 2: Lexicographical Selection}
        \STATE $i^* \gets \text{null}$
        \FOR{$n \in \mathcal{N}_{safe}$}
            \IF{$i^* = \text{null} \textbf{ or } \text{LexCompare}(n, i^*)$}
                \STATE $i^* \gets n$
            \ENDIF
        \ENDFOR

        \STATE \textbf{Step 3: Assignment \& state Update}
        \STATE Assign $r$ to DP unit $i^*$
        \STATE $B_{i^*} \gets B_{i^*} + 1$
        \STATE $K_{i^*} \gets K_{i^*} + \text{Length}(r)$
    \ENDFOR
\ENDFUNCTION
\end{algorithmic}
\end{algorithm}

\subsubsection{Outlier-Resilient Load Balancing}

The distribution of KVCache lengths in conversational workloads is typically heavy-tailed, rendering mean-variance based metrics unstable. To robustly identify straggler DP units without over-sensitivity to normal variance, we employ the Interquartile Range (IQR) method.

By defining a dynamic exclusion threshold $Th_{outlier} = Q_3 + k \cdot IQR$, the scheduler effectively masks DP units at risk of memory exhaustion or computational saturation. This "Mask-then-Select" approach creates a safe decision space for the subsequent lexicographical optimization of batch size and compute load.

To maximize the efficacy of this masking strategy, we implement \textbf{Length-Based Pre-Sorting}: requests within a batch are sorted in descending order of total sequence length. This structure facilitates a "fill-the-valley" placement strategy, prioritizing heavy requests while the decision space is still abundant.

\subsubsection{Lexicographical Multi-Objective Selection}

For the DP units remaining in the "safe decision space" (post-masking), we model the assignment as a \textbf{Lexicographical Optimization} problem. We define the state vector for DP unit $i$ as $V_i = \langle B_i, K_i \rangle$, where $B_i$ is the batch size and $K_i$ is the KV cache length. The optimal DP unit $i^*$ is selected by hierarchical minimization:

$$i^* = \arg\min_{i \in \mathcal{D}_{valid}} \langle B_i, K_i \rangle$$

This logic prioritizes balancing the Batch Size ($B_i$) to maximize parallel efficiency, while using KV Cache load ($K_i$) as a tie-breaker to manage memory pressure. This hierarchical approach ensures the cluster converges towards an equilibrium where both compute utilization and memory footprint are optimized.

\section{Experiments}
\label{sec:experiments}

\subsection{TTFT Optimization}

\textbf{Experimental Setup:} We evaluated the proposed Staggered Batch Scheduling (SBS) on a production cluster equipped with NVIDIA H800 GPUs. The cluster topology followed a 3:1 Prefill-to-Decode (3P1D) ratio, utilizing the DeepSeek-V3 model. Each inference instance was configured with Prefill Chunk Size (3K), Tensor Parallelism ($TP=4$), Data Parallelism ($DP=8$), and Expert Parallelism ($EP=32$). The evaluation workload comprised requests with input token lengths ranging from 0 to 3K (mean: 1K).

\begin{figure}[h]
    \centering
    \subfigure[\textbf{Input Token Length(0-3K):} SBS maintains a consistent TTFT advantage across all load levels.]{
        \includegraphics[width=0.36\textwidth]{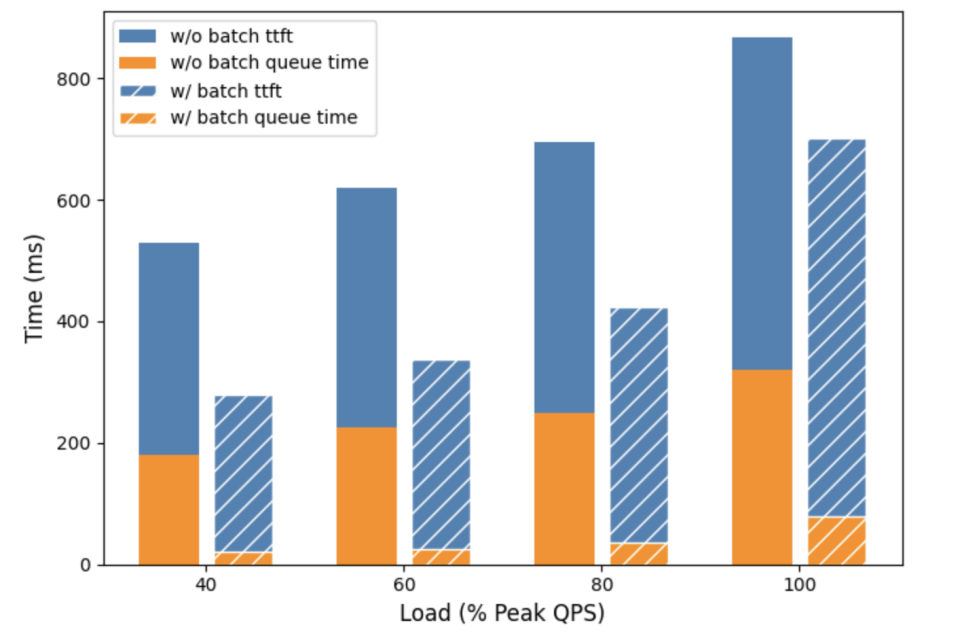}
        \label{fig:prefill_0_3k}
    }
    \subfigure[\textbf{Input Token Length(3K-64K):} Under high-variance inputs (up to 64K tokens), SBS effectively suppresses tail latency, demonstrating robustness in complex production scenarios.]{
        \includegraphics[width=0.36\textwidth]{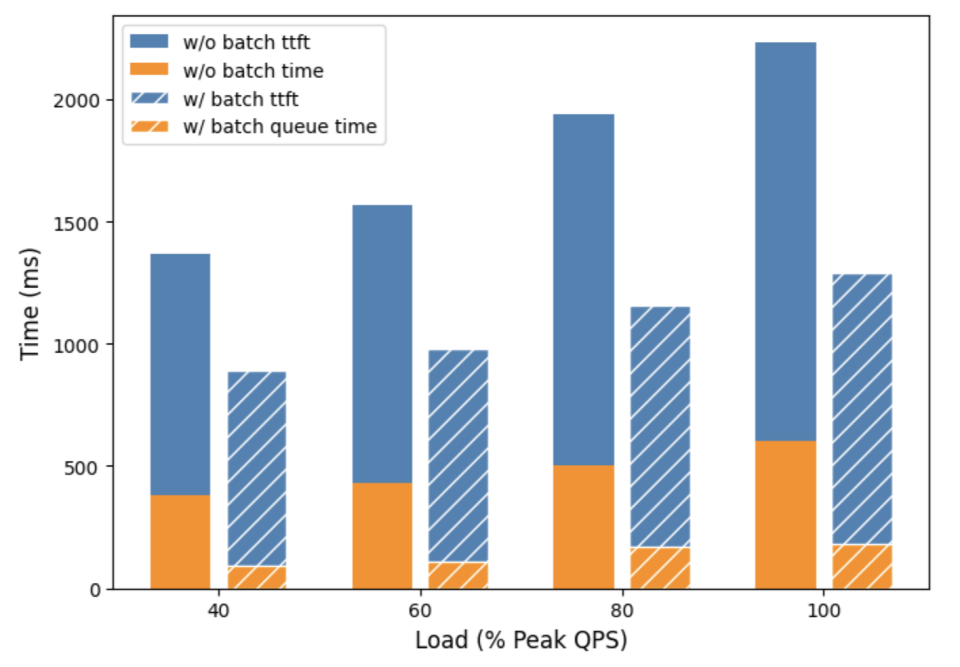}
        \label{fig:prefill_3_64k}
    }
    \caption{Prefill Performance Evaluation}
\end{figure}

We compared the average Time-to-First-Token (TTFT) and internal queuing latency of SBS against a baseline immediate-dispatch scheduler. To establish a reference capacity, we first benchmarked the baseline to determine its peak QPS that satisfies the TTFT Service Level Objective (SLO). Subsequently, we evaluated both systems under \textbf{identical QPS conditions} across load levels ranging from 40\% to 100\% of this peak. As shown in Figure \ref{fig:prefill_0_3k}, SBS consistently outperforms the baseline. By substantially reducing internal queuing delay, SBS achieves up to a \textbf{40\% reduction in TTFT at sub-80\% load levels}.

We further extended the evaluation to long-context scenarios (3K-64K input tokens, mean 6.7K) using a configuration with a 16K Prefill Chunk Size. Figure \ref{fig:prefill_3_64k} demonstrates that SBS maintains its performance advantage even under high variance in input lengths, validating its robustness for complex, production-grade workloads.

\subsection{Load Balancing Effects}

\subsubsection{Prefill Load Balancing and Throughput}

To quantify the efficacy of the "Water-Filling" allocation strategy (Algorithm \ref{alg:pbaa}), we conducted controlled experiments comparing Baseline and SBS modes under \textbf{fixed mean-TTFT constraints}. 

Table \ref{tab:batch_scheduling_performance} details the maximum sustainable QPS and average Prefill Chunk Utilization—a metric quantifying the percentage of theoretical token capacity utilized per forward pass. The results demonstrate that SBS effectively converts fragmented "parallelization bubbles" into usable throughput. By employing aggressive batching and bin-packing, \textbf{Prefill Chunk Utilization surged from ~52\% to ~88\%}. This substantial gain in resource efficiency translates directly into system capacity, boosting \textbf{overall QPS by 12.9\% to 22.8\%} across different prefill chunk size configurations. These findings confirm the efficacy of SBS in mitigating load imbalance across DP units.

\begin{table*}[h]
\centering
\caption{Prefill Chunk Utilization and Total System Throughput Comparison.}
\label{tab:batch_scheduling_performance}
\begin{tabular}{@{}lcccccc@{}}
\toprule
\textbf{Scenario} & \textbf{Batch} & \textbf{QPS} & \textbf{Chunk Util. (\%)} & \textbf{$\Delta$QPS (\%)} & \textbf{$\Delta$ Chunk Util. (pp)} \\
\midrule
\multirow{2}{*}{Chunk 3K (mean-TTFT=0.8s)} 
& Off & 57 & 51.83 & -- & -- \\
& On & 70 & 88.7 & +22.8 & +36.9 \\
\cmidrule(r){1-6}
\multirow{2}{*}{Chunk 5K (mean-TTFT=1.0s)}
& Off & 70 & 53.0 & -- & -- \\
& On & 79 & 88.0 & +12.9 & +35.0 \\
\bottomrule
\end{tabular}
\caption*{\footnotesize By eliminating parallelization bubbles via batched bin-packing, SBS increases Prefill Chunk Utilization from $\sim$52\% to $\sim$88\%. This improved resource saturation directly drives a 12.9\% to 22.8\% boost in maximum sustainable QPS compared to the baseline.}
\end{table*}

\subsubsection{Decode Load Balancing and Throughput}
We evaluated the Decode phase on an H800 cluster configured with TP=1, DP=32 and EP=32, handling a representative workload with combined input and output lengths of approximately 2.5K tokens and an average batch size of 35. We compared our \textbf{IQR-Aware Lexicographical Scheduling} against a standard immediate-dispatch baseline.

\begin{figure}[h]
    \centering
    \includegraphics[width=0.45\textwidth]{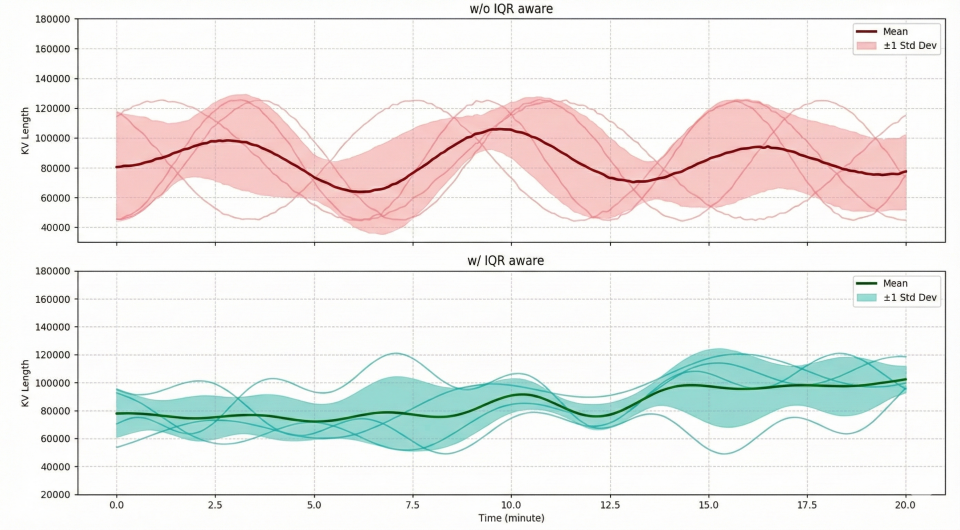}
    \caption{\textbf{Efficacy of IQR-Aware Load Balancing on Decode Phase.}
The visualization compares the distribution of KV Cache loads across DP units over time.
\textbf{Top (Baseline):} Standard scheduling results in a heavy-tailed distribution with wide variance (red band), indicating frequent stragglers.
\textbf{Bottom (SBS):} Our IQR-Aware strategy compresses the load variance (green band), keeping KV Cache usage tightly clustered around the mean. This reduction in load disparity ($\pm 1\sigma$ range reduced by ~40\%) directly mitigates the synchronization overhead.}
    \label{fig:decode_balance}
\end{figure}

\textbf{Straggler Suppression:} As shown in Figure \ref{fig:decode_balance}, the baseline strategy resulted in a heavy-tailed distribution of KV Cache loads, with a standard deviation band ($\pm 1\sigma$) spanning \textbf{40k to 130k tokens}, and peak outliers approaching 150k. In contrast, our IQR-Aware approach effectively compressed the load variance, stabilizing the $\pm 1\sigma$ range between \textbf{60k and 100k tokens} and eliminating extreme outliers.

\begin{figure}[h]
    \centering
    \includegraphics[width=0.45\textwidth]{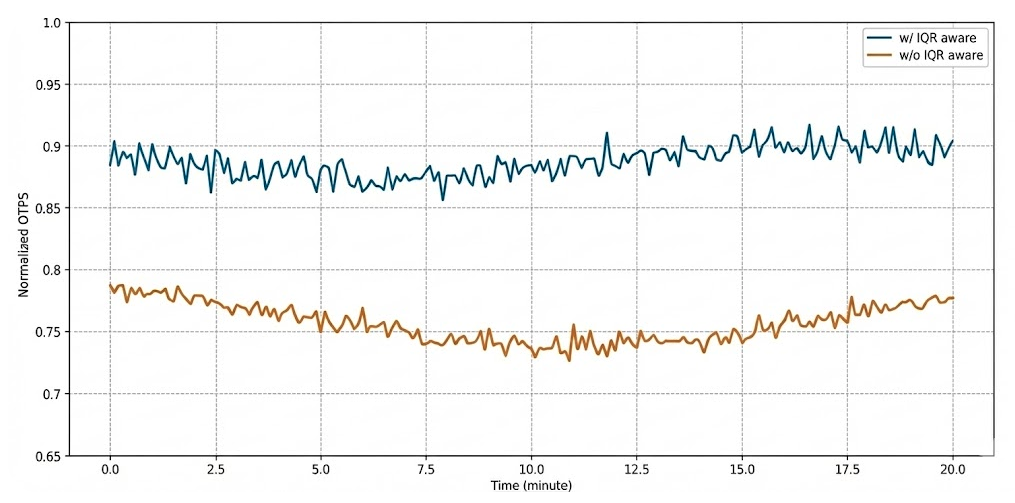}
    \caption{\textbf{Impact of Load Balancing on Decode Throughput.} By effectively suppressing straggler DP units (as evidenced in Figure \ref{fig:decode_balance}), the SBS strategy minimizes idle wait times caused by synchronization barriers. This optimization converts previously wasted parallelization bubbles into productive token generation, resulting in a 15\% increase in aggregate Decode throughput.}
    \label{fig:decode_throughput}
\end{figure}

\textbf{Throughput Gain:} The improved load balance directly reduced the "straggler effect" in the synchronization barrier. Consequently, the aggregated \textbf{Decode Throughput increased by 15\%}. These results validate that jointly optimizing the coupled dimensions of compute load and memory footprint directly translates into superior system capacity and resource utilization. 

\section{Conclusion}
\label{sec:conclusion}
In this work, we addressed the scheduling inefficiencies inherent in P/D-separated, large-scale DP+EP inference architectures. We demonstrated that the \textbf{coupling of immediate dispatch strategies with high-synchronization parallel execution} is the root cause of latency degradation.

Our proposed solution, \textbf{Staggered Batch Scheduling (SBS)}, fundamentally shifts the scheduling paradigm from continuous dispatch to \textbf{discrete, window-based allocation}. By proactively buffering requests, SBS eliminates internal queuing latencies and parallelization bubbles, \textbf{reducing TTFT by up to 40\%}. Furthermore, Leveraging the batching window, we deploy a global allocation mechanism across both phases. It \textbf{boosts Prefill throughput by 12.9\%–22.8\%} via bin-packing, and \textbf{improves Decode throughput by $\sim$15\%} by resolving the coupled challenges of KV-cache variance and batch size imbalance. 

As LLMs continue to scale towards trillion-parameter mixtures of experts, the principles of \textbf{discrete scheduling synchronization} and \textbf{global-view load shaping} presented here provide a scalable path for next-generation inference systems.

\newpage


\bibliography{references}
\bibliographystyle{mlsys2025}


\end{document}